\documentclass{article}

\usepackage{graphicx,amsmath,amssymb,feynmf,dsfont,latexsym,braket,pictexwd}
\usepackage{undertilde}
\usepackage[usenames,dvipsnames]{color}
\definecolor{blue}{rgb}{0,0,.7}
\definecolor{red}{rgb}{.7,0,0}
\definecolor{orange}{rgb}{1,.6,0}
\definecolor{purple}{rgb}{.4,0,.5}
\definecolor{brown}{rgb}{.4,.2,.1}
\definecolor{green}{rgb}{0,.57,0}
\usepackage[margin=2cm]{geometry} 
 
\newcommand{\cO}{\mathcal{O}}
\newcommand{\cE}{\mathcal{E}}

\newcommand{\bi}{\begin{itemize}}
\newcommand{\ei}{\end{itemize}}

\newcommand{\del}{\nabla}
\newcommand{\ben}{\begin{enumerate}}
\newcommand{\een}{\end{enumerate}}

\newcommand{\ds}{\displaystyle}

\newcommand{\cL}{\mathcal{L}}
\newcommand{\cA}{\mathcal{A}}

\def\dot{\!\cdot\!}

\def\bea{\begin{eqnarray}}
\def\eea{\end{eqnarray}}

\usepackage[normalem]{ulem}
\usepackage{verbatim}
\usepackage{makeidx}
\makeindex

\renewcommand{\author}[1]{\vspace{2ex}{\large\begin{center}
 \setlength{\baselineskip}{3ex}#1\par\end{center}}}
\renewcommand{\thanks}[1]{\footnote{#1}}

\usepackage{bbm}
\newcommand{\untilde}[1]{\mathbbm{{#1}}}
\begin{document}

\centerline{\sc \large  Towards an Effective Field Theory on the Light-Shell}
\author{
 Howard~Georgi,\thanks{\noindent \tt georgi@physics.harvard.edu}
Greg~Kestin,\thanks{\noindent \tt kestin@physics.harvard.edu} 
Aqil~Sajjad,\thanks{\noindent \tt sajjad@physics.harvard.edu} 
 \\ \medskip
Center for the Fundamental Laws of Nature\\
Jefferson Physical Laboratory \\
Harvard University \\
Cambridge, MA 02138
 }
\begin{abstract}
We discuss our work toward the construction of a light-shell effective theory (LSET), an
effective field theory for describing the matter emerging from high-energy
collisions and the accompanying radiation. We work in the highly simplified
venue of 0-flavor scalar quantum electrodynamics, with a gauge invariant
product of scalar fields at the origin of space-time as the source of
high-energy charged particles. Working in this simple gauge theory allows
us to focus on the essential features of LSET. 
We describe how the effective theory is constructed and argue that it can
reproduce the full theory tree-level amplitude.  We study the 1-loop
radiative corrections in the LSET and suggest how the leading
double-logs in the full theory at 1-loop order can be reproduced by a
purely angular integral in the LSET. 
\end{abstract} 
\newpage
\section{Introduction}

Previously, \cite{Georgi:2010nq} we expressed the hope of constructing 
an effective field theory on the
2-dimensional light-shell emerging from a high-energy collision. 
The idea was motivated by a classical picture
of a very high energy hadronic collision in which colored particles are
produced from an initially color-neutral state at $t=r=0$
and instantaneously accelerate outward to the speed of light. At the same time, the collision
event produces a pulse of color radiation that also moves out at the speed
of light.
 So, classically, in the very high energy limit, everything lies
on an expanding sphere at $t=r$, which we call the light-shell. It is worth mentioning that some of the recent work on assymptotic gauge symmetries has been exploring some related themes involving the null sphere at infinity \cite{He:2015zea, Pasterski:2015zua, Adamo:2014yya, Adamo:2015fwa, Campiglia:2015qka}.

In section \ref{background} the classical calculation is discussed in more
detail. There we observe that to have the 
vector potential
$A^\mu$ confined to the light-shell  requires a special gauge - $v_\mu
A^\mu=0,$ where 
$v^\mu$ is a light-like 
vector pointing away from the origin. We call this light-shell gauge (LSG).  

In this paper, we begin the explicit construction of the effective field
theory, incorporating the intuition gained from the classical
picture by studying the quantum mechanics of particle production
  from a gauge invariant source at the origin of space-time. We
will see how a gauge invariant
product of scalar fields at the origin of space-time 
gives rise to an effective field
theory of the high energy physics that depends only on the angles of the
momenta of the high energy particles and fields. This
2-dimensional effective theory is our 
light-shell effective theory (LSET). 
Here we present it in the simplified venue of 0-flavor scalar quantum
electrodynamics (sQED). 
This strips away most of the physics so that we can focus on
the construction of the LSET.   

The reader might wonder why we are investing so much effort in a new
effective theory so novel that we have to restrict ourselves to studying it
in a toy model when there is a well-developed theory, SCET
\cite{Bauer:2000ew,Bauer:2000yr} that is already being successfully applied
to QCD processes at high energies \cite{Schwartz:2007ib}. The real answer,
of course, is that we find it fascinating because our approach is very
different. So, while we aim eventually to generalize our construction to
QCD, even before then we hope that insight can be taken from the picture we
begin to describe here. For example, an interesting feature we will
discuss is that all our calculations can all be reduced to purely angular
integrals.     

In the classical picture, the starting point is a very high energy event in
which charged particles are produced at the origin.
Appropriately translating this classical setup
into quantum field theory suggests a gauge invariant source at $t=r=0$.
All of the physics in the effective theory will come from such a source
term in our LSET Lagrangian. Since LSG is undefined at $r=0$, we construct the
source on a small sphere around the origin whose size 
we \textit{eventually} shrink to zero. In this introduction, we will also
be ignoring the initial state and focusing just on the physics coming from
our source. In leading order, the form of the source is fixed by gauge invariance;
in section \ref{EFTsource} we obtain the explicit form of the source. 

The EFT requires that we set an energy scale $\cE$ to define what we mean
by ``high energy''.  In the spirit of HQET \cite{Georgi:1990um} (for a
recent and comprehensive review see \cite{Mannel:2004ce})  we  scale
out the  large momenta associated with the energetic outgoing
particles. 
The associated decomposition into fields above and below $\cE$ we call the
large radial energy (LRE) expansion, and the related fields are termed LRE
fields. The LRE fields correspond to high-energy particles produced by the
source carrying large energy $E>\cE$ outwards 
from $t=r=0$ into the bulk space. We will see that to leading order in
$1/E$, the direction $\hat r$ of propagation away from the origin
is a classical variable and we can
label the LRE fields by $\hat r$.  But in the presence of interactions, 
the directions of the LRE fields cannot be specified precisely.
So to each charged LRE field we assign 
an ``angular size''. 

To obtain the LSET Lagrangian, we apply the LRE expansion to sQED, and
expand in orders of $1/E$. We note that the gauge interactions at leading
order (in $1/E$) 
are proportional to $v^\mu A_\mu$. So choosing LSG eliminates the gauge
interactions of the LRE fields, 
showing that these interactions are just gauge artifacts, and that as
expected, all the physics at high energies is occurring at the origin and depends
only on angles. This suggests that a source at the origin is sufficient to
describe the physics in LSET. Because different configurations of LRE
fields (different 
energies and directions) do not interfere, each such configuration is
associated with its own sector, and the source in the EFT is a sum over all
such sectors, separated by superselection rules. 

As with any EFT, one must match to the full theory. In section
\ref{scalarmatching} the LRE fields are matched to the full theory
fields.  Then in section \ref{treelevel} we confirm the theory's
structure by showing that the source in the EFT 
reproduces the tree-level amplitude for the source in
the full theory in the appropriate limit. 

Finally, we calculate the LSET
1-loop correction for production of a scalar/anti-scalar pair.  The
relevant photon propagator is the LSG bulk propagator in the limit that the
end points go to the light shell.  
We will show that for
physically sensible relations between the cut-offs and $\cE$, the
effective field theory can reproduce
the leading double-logs of 0-flavor sQED.   

\section{Background}
\label{background}

LSET is motivated by a simple classical electrodynamics
calculation. Consider N charged particles  
created at 
$\vec r=0$ and $t=0$, which then instantaneously accelerate to the
speed of light with velocities $c\hat 
r_{j}.$ The resulting electric and magnetic
fields are zero everywhere in space except on an infinitesimally thin
spherical shell that expands out from the origin at the speed of light. In \cite{Georgi:2010nq}, we not only motivated this idea in the context of electromagnetism but also showed that a similar picture holds for the classical non-abelian theory with the color fields associated with a non-linear $\sigma$-model on the 2D light-shell with specific symmetry breaking terms. In this paper, however, we will limit our discussion to the abelian case.

The starting point for our quantum mechanical description of this picture is a gauge
  invariant source at the origin of space-time that produces charged
  particles moving away from the origin. Gauge invariance of the source
  requires that the total charge vanishes, 
\begin{equation}
\sum_j\,q_j=0
\end{equation}  Because we consider a point
  source at the origin, our source does not conserve energy and momentum
  and can produce final states with any physical energy and momentum.  We
  focus on the final states with some large energy of order $E$.

The classical observation motivates the picture of LSET, while also motivating the
gauge in which we will work. If one makes an appropriate gauge
transformation, not only will the electric and magnetic fields lie on a
spherical shell, but the potential will as well. For the general charge
configuration described above, in the appropriate gauge, our potentials are
\begin{equation}
A^0\left(t,\vec r\,\right)=\sum_j^N q_j\,\delta(t-r)\,\log\left(t+\hat r_j\cdot\vec r\right)
\end{equation}
\begin{equation}
\vec A\left(t,\vec r\,\right)
=\,\hat r\,\sum_j^N q_j\,\delta(t-r)\,\log\left(t+\hat r_j\cdot\vec
  r\right)
\label{classicalpotential}
\end{equation}
The potentials are proportional to a $\delta$-function that sets $t=r$, so
they lie on the ``light-shell". An important feature is that we can write
$\vec A\left(t,\vec r\,\right) =\hat r\,A^0\left(t,\vec r\,\right)$ 
satisfying the constraint 
\begin{equation} v_\mu
A^\mu=0
\quad\mbox{where}\quad
v^0=1\quad\mbox{and}\quad\vec v=\hat r
\label{v-def}
\end{equation}
The vector $v^\mu$
is a light-like vector {\bf that varies with position.} 
(\ref{v-def}) defines light-shell
gauge (LSG), an essential component of the LSET
effective theory. 

Moving to a theory involving quantum effects, we discuss the formulation in the
simplest theory, 
0-flavor scalar quantum electrodynamics. In the subsequent sections, we
will combine sQED with new ingredients in order to construct the the
effective theory. Before doing so, we write down the important pieces of the
full theory Lagrangian\footnote{We do not write the scalar
  self-interactions because they play no role in the limit we consider.}
in order to refer to it later on as we mix in these new elements.  
\begin{equation}
\mathcal{L}_{\rm full}
=\underbrace{\left|\partial^\mu\phi\right|^2}_{\mathcal{L}_{\rm \phi}}
+\underbrace{\left(e^2\,A^\mu\phi\,A^\mu\phi^*-ieA^\mu((\partial^\mu\phi)\,\phi^*
+(\partial^\mu\phi^*)\phi )\right)}_{\mathcal{L}_{\rm int}}
\underbrace{-\frac{1}{4}F_{\mu\nu}^2}_{\mathcal{L}_{\rm A}}
\label{Lfull}
\end{equation}
and we are interested in the functional integral
\begin{equation}
\left.\int\,S_{\rm full}\,e^{i\int\mathcal{L}_{\rm full}\,d^4x}\,[d\phi]\,[dA]
\right/
\int\,e^{i\int\mathcal{L}_{\rm full}\,d^4x}\,[d\phi]\,[dA]
\label{functional}
\end{equation}
Where $S_{\rm full}$ is whatever gauge invariant source at the origin we
wish to consider. The source describes the fields which are produced at the
origin; so for example if we wish to consider the creation of a
scalar/anti-scalar pair, then $\ds S_{\rm full}=\phi^*(0)\phi(0).$  

The restriction to ``0-flavor''
allows us to ignore matter loops and (at least at one loop) the
  scalar self-couplings (for details see Appendix \ref{0-flavor}) in our radiative corrections, which makes
the physics simpler and allows us to focus narrowly on the
construction of the effective field theory. Similarly, the restriction to
``scalar'' QED allows us to simplify the discussion of the matter in the theory.
So, in theories of this kind, the only important physics is associated with
the source and the radiation generated by it, and we can focus on the
construction of the novel effective theory. The details of the source in
LSET will be described in section \ref{EFTsource}. 

\section{Constructing the light-shell effective theory Lagrangian}
\label{construct}

In this section we construct the terms of the LSET Lagrangian that
correspond to the terms in equation (\ref{Lfull}), $\mathcal{L}_{\phi}$,
$\mathcal{L}_{int}$, $\mathcal{L}_{A}$, and the source. As with any
effective theory, the soft physics is left unchanged, so we focus on the
physics associated with hard particles, and we distinguish the part of the
LSET Lagrangian involving hard particles by referring to it as
$\mathcal{L}_{LRE}$. This construction involves two new and essential
ingredients: a field decomposition we will refer to as a large radial
energy (LRE) expansion and light-shell gauge.   

The large radial energy expansion is reminiscent of the field
decomposition of 
HQET \cite{Georgi:1990um} and LEET (the precursor of SCET
\cite{Dugan:1990de} that sums soft logs but not collinear logs). We scale out the
uninteresting large 
momenta associated with the 
energetic particles, but as its name suggests the LRE expansion involves
scaling out by a spherical wave. In order to do this, we set an energy scale
$\cE$, that determines which fields are large radial energy fields
$\Phi_E^{(*)}$, and which fields are soft $\phi_s$. The decomposition is
{\renewcommand{\arraystretch}{2.5}
\begin{equation}
\begin{array}{c}
\displaystyle
\phi=\phi_s+\sum_{E>\cE}\left( \frac{e^{-iE(t-r)}}{\sqrt{2E}}\Phi_{E,+q}
+\frac{e^{iE(t-r)}}{\sqrt{2E}\;}\Phi_{E,-q}^* \right)
\\ \displaystyle
\phi^*=\phi_s^*+\sum_{E>\cE}\left( \frac{e^{-iE(t-r)}}{\sqrt{2E}}\Phi_{E,-q}
+\frac{e^{iE(t-r)}}{\sqrt{2E}\;}\Phi_{E,+q}^* \right)
\end{array}
\label{LREscalars}
\end{equation}
}where $\Phi_{E,\pm q}$ ( $\Phi_{E,\pm q}^*$) annihilates (creates) high
energy outgoing scalars with charge $\pm q$. In the following, we will
focus on the particles with charge $+q$ and drop the $\pm q$ subscripts to
simplify the tableaux. 
As usual in such an effective field theory decomposition, the $x$
dependence of the EFT field is assumed to be slow compared to the
$t$ and $r$ dependence of the exponential factor $e^{iE(t-r)}$, and
derivatives of $\Phi_E$ are assumed to be small compared to $\cE$ in
the effective theory.\footnote{This is a bit trickier than it sounds.  See
 \protect\cite{Luke:1992cs}.} The $1/\sqrt{2E}$ is a normalization, the reason for
which will soon be apparent. 

Applying this expansion, the LSET\ Lagrangian can be written as an
expansion in the small parameter $(1/E)$, where $E$ is the energy scaled out
of the energetic field at hand. Let's begin to look at $\cL_{LRE}$ by
examining $\mathcal{L}_{\rm \phi}$, the kinetic energy of our matter field,
to leading order in $1/E$. Using our expansion (\ref{LREscalars}), focusing on the
LRE terms, and using
\begin{equation}
\frac{\partial(t-r)}{\partial x_\mu}=v^\mu
\end{equation}
we get  
\begin{equation}
(D^\mu\phi)^*\,D_\mu\phi\to
\frac{1}{2E}
\,\Bigl((D^\mu+iEv^\mu)\Phi^*_E\Bigr)
\,\Bigl((D^\mu-iEv^\mu)\Phi_E\Bigr)
\label{lrelagr}
\end{equation}
The cross terms are leading in the $1/E$ expansion, and have a factor of
$E$ from the derivatives acting on the spherical wave, which cancels the
normalization from (\ref{LREscalars}), giving 
\begin{equation}
=i\,\Phi_E^*\,\Bigl(
\partial_t+(\hat r\cdot\vec\nabla+\vec\nabla\cdot\hat r)/2
\Bigr)\Phi_E
+\frac{1}{2Er^2}\Phi_E^*\,\tilde L^2\,
\Phi_E
\label{lreke}
\end{equation}
where the $\ds
\tilde L^2=
-r^2\,(\vec\nabla_\perp^T-iq\vec A_\perp)\cdot(\vec\nabla_\perp-iq\vec A_\perp)
$ and we omit terms that vanish by the zeroth-order equations of motion.
While the $\tilde L^2$ term is of order $1/E$, it also has rapid $r$
dependence as $r\to0$, which we do not want.
We can make the following field redefinition to eliminate it:
\begin{equation}
\tilde \Phi_E(x)\equiv \exp\left[i\frac{\tilde L^2}{2Er}\right]\,\Phi_E(x)
\end{equation}
Note that since the derivatives in $\tilde L^2$ are all covariant,
$\tilde\Phi_E(x)$ transforms just like $\Phi_E(x)$ under gauge
transformations. In terms of $\tilde \Phi_E(x)$, and ignoring interaction
terms, the kinetic energy becomes 
\begin{equation}
\mathcal{L}_{\rm \phi}
=i\,\tilde\Phi_E^*\,\Bigl((\partial_\mu v^\mu+v^\mu\partial_\mu)/2
\Bigr)\,\tilde\Phi_E
=i\,\tilde\Phi_E^*\,\Bigl(
\partial_t+(\hat r\cdot\vec\nabla+\vec\nabla\cdot\hat r)/2
\Bigr)\,\tilde\Phi_E
\label{lreket2}
\end{equation}
The kinetic energy term (\ref{lreket2}) looks very much like the
corresponding terms in HQET \cite{Georgi:1990um}
and LEET \cite{Dugan:1990de}, but there the analog of the vector
$v^\mu$ is a constant, time-like in HQET and light-like in LEET. The fact
that $v^\mu$ varies with $\hat r$ is responsible for unique properties
of the LRE expansion. For example, the LRE decomposition (\ref{LREscalars})
is invariant under rotations about the origin, not just covariant like HQET
or LEET.

The $\tilde\Phi_E$ propagator associated with the
  kinetic energy term (\ref{lreket2}) is directional and has the
form\footnote{When $\hat 
  r$ appears as an argument, it refers to dependence on angles $\theta$ and
  $\phi$. Likewise $\hat r_j$ refers to the angles $\theta_j$ and
  $\phi_j$. So, here $\delta(\hat r-\hat r')$ is equal to
  $\delta(z-z')\delta(\phi-\phi')$, with $z=\cos(\theta)$.} 
\begin{equation}
\Braket{0|T\,\tilde\Phi_E(x)\,\tilde\Phi_E^*(x')|0}
=\frac{1}{rr'}\,\theta(t-t')\,\delta(t-r-t'+r')\,
\delta(\hat r-\hat r')
\label{lreprop0}
\end{equation}
One can check (\ref{lreprop0}) easily and it can be formally derived using
canonical quantization, as we show in appendix \ref{canquant}. The propagator
(\ref{lreprop0}) describes radially outgoing particles and this
  form establishes the connection between the spatial direction 
of the coordinate $x$ and the direction of propagation of the particle,
which determines the direction of the momentum of the LRE particle far away
from the source. This connection between position space and momentum space
for the high-energy particles is one of the crucial components of our
construction.  We will return to this and see the connection very
  explicitly in section \ref{scalarmatching}. But while the connection is
exact in the free theory, we 
would expect that quantum 
effects make it impossible to specify the momentum direction precisely. This
expectation is reified in the calculation of quantum loops where specifying
the directions precisely leads to divergences \cite{LSETrunning}. We assume
that this is 
associated with the physical impossibility of measuring a jet direction
exactly. Thus we associate an angular size with each LRE particle quantifying
the uncertainty in direction.  

We now return to  equation (\ref{lrelagr}) to explore the consequences of the LRE expansion for  $\mathcal{L}_{\rm int}$. It can be written in the suggestive form
\begin{equation}
\frac{i}{2} 
\Bigl[(-\partial^\mu \Phi_E^\dagger) v_\mu \Phi_E +v^\mu \Phi_E^\dagger \partial_\mu \Phi_E\Bigr]
\,+\, v_\mu A^\mu\Phi_E^\dagger \Phi_E
\,+\,\frac{1}{2E} (D^\mu\Phi_E)^\dagger D_\mu\Phi_E
\label{suggestive}
\end{equation}
In this form, it is clear that in LSG our interactions vanish at leading
order.   The removal of the
gauge interactions with LRE scalars simplifies calculations, and it 
makes it clear that the essential physics of the high-energy particles
is associated with
the source at the origin. This is consistent with the expectation of a
purely angular theory on the light-shell. 

Lastly, consider the purely gauge part of the Lagrangian. In LSG,
$\mathcal{L}_{\rm A}$ becomes\footnote{This form is not completely obvious,
  at least to us.  Details of the derivation can
be found in \cite{LSGprop}.} 
\begin{equation}
\cL_{A}=-\ds\frac{1}{4}F_{\mu\nu}^2=-\frac{1}{2}\biggl(A_r,\vec A^T_\perp\biggr)
\,\begin{pmatrix}
(\partial_t+\vec\nabla\cdot\hat r)
(\partial_t+\hat r\cdot\vec\nabla)&(\partial_t+\vec\nabla\cdot\hat r)\vec\nabla^T \\
\vec\nabla
(\partial_t+\hat r\cdot\vec\nabla)& 
\vec\nabla\vec\nabla^T+\Box\,I \\
\end{pmatrix}\,\begin{pmatrix}
A_r  \\
\vec A_\perp  \\
\end{pmatrix}
\label{LagA}
\end{equation}
Here we have used $A_r$ and $\vec A_\perp$, defined by
\begin{equation}
A_r=\hat r\cdot\vec A
\end{equation}
\begin{equation}
\vec A_\perp=\vec A-\hat r\cdot\vec A \,\hat r
\end{equation}
Keep in mind that in LSG the temporal component of $A^\mu$ is equal to $A_r$. An LRE expansion, similar to that of the scalars, holds for the gauge
field. This expansion is in terms of longitudinal and perpendicular
components of the gauge field, which is appropriate in light-shell
gauge. Again, the rescaling of each field is determined by the canonical form of the kinetic energy term.
\begin{equation}
\vec A =\vec A_s+\sum_{E>\cE}\left(\frac{1}{\sqrt{2E}}\,e^{iE(t-r)}\,\vec\cA^*_{E\perp}+
\frac{1}{\sqrt{2E}}\,e^{-iE(t-r)}\,\vec\cA_{E\perp}+
e^{iE(t-r)}\,\cA^*_{Er}\hat r+
e^{-iE(t-r)}\,\cA_{Er}\hat r+\cdots\right)
\label{LSGLRExp}
\end{equation}
After applying this expansion to (\ref{LagA}) and considering  the LRE
terms,  we can redefine the gauge field as 
\begin{equation}
\begin{pmatrix}
\untilde{A}_{Er}\\ \vec{\untilde{A}}_{E\perp}
\end{pmatrix}=\begin{pmatrix}
1&
(\partial_t+\hat R\cdot\vec\nabla)^{-1}\,\vec\nabla^T/\sqrt{2E} \cr
0&1\cr
\end{pmatrix}
\begin{pmatrix}
A_{Er}\\ \vec A_{E\perp}
\end{pmatrix}
\label{DiagTrans}
\end{equation}
Where $(\partial_t +\vec\nabla\dot\hat R)^{-1}$ is the inverse
  of a differential operator that is non-local in space and time and given explicitly by
\begin{equation}
(\partial_t +\vec\nabla\dot\hat R)^{-1}(x',x)=\frac{1}{(r')^2}\theta(t'-t)\delta(t'-r'-t+r)\delta(\hat r'-\hat r)
\label{dtdrinverse}
\end{equation}
 Operators of this sort appear frequently in our LSET analysis (though we will spare the
  reader of this introduction most of the gory details --- these will
  appear in \cite{LSGprop} and \cite{LSETrunning}).   These operators are treated on the same footing as linear operators, so for example, the first row of (\ref{DiagTrans})
could be written more explicitly as
\begin{equation}
\untilde A_{Er}(x)=A_{Er}(x)+\int
\,\left((\partial_t+\hat R\cdot\vec\nabla)^{-1}(x,x')
\,\vec\nabla^{'T}/\sqrt{2E}\,\vec A_{E\perp}(x')\,d^4x'\right)
\label{DiagTransExplicit}
\end{equation}
In section (\ref{treelevel}) we will see that the quanta of the
$\vec{\untilde A}_\perp$ field can be
directly related to those of the full 
theory. Also, this  allows us to write the LRE\ photon kinetic energy in
the following diagonal form. 
\begin{equation}
\cL_{A,\,LRE}=
\begin{pmatrix}
\untilde A^*_{Er}&
\vec{\untilde{A}}^*_{E\perp} \cr
\end{pmatrix}
\begin{pmatrix}
-\,
(\partial_t+\vec\nabla\cdot\hat r)(\partial_t+\hat r\cdot\vec\nabla)& 0\cr
0&(\partial_t+\hat r\cdot\vec\nabla/2+\vec\nabla\cdot\hat r/2)  \cr
\end{pmatrix}
\begin{pmatrix}
\untilde A_{Er}\cr
\vec{\untilde{A}}_{E\perp} \cr
\end{pmatrix}
\end{equation}The final piece of the LSET Lagrangian is the source. This is
where the physics of our theory lies, and we describe
  it in the following section. 

\section{LSET source}
\label{EFTsource}
So far we have constructed the LSET Lagrangian by bringing the large radial
energy expansion and light-shell gauge to the full theory. In doing so we
have removed all of the interactions of the LRE particles 
except for those directly associated
with the point source at the origin in the full theory. 
The full-theory source is proportional to a gauge invariant
product of local fields at the origin. Thus we also expect the corresponding
source in the EFT to be gauge invariant.
The
conventions for the gauge transformations of our fields are listed in
appendix \ref{conventions}. 

While the full-theory source is at the origin, light-shell gauge is
ill-defined there, so we begin by considering a source in the EFT that is ``spread
out"  about the origin. We also expect from our classical
picture that as the energy in the 
  event goes to infinity, all of the physics goes onto the light shell, at
$t=r$.  Thus in our quantum version,
we spread out our source onto a surface $r=s$
surrounding the origin, near the light-shell, with $t-r\to0$ as $E\to\infty$.
To understand the symmetry of the spread-out source, it
is convenient to write $\varphi(x)=\varphi(t,r,\hat r)$ and to let
\begin{equation}
\varphi(r,r,\hat r)\equiv \left.\varphi(x)\right|_{t=r}
\end{equation}
represent either an LRE field or a soft field on the light shell. 
When we eventually write
down the full source, the LRE fields will be evaluated at particular values
of $\hat r$, while the soft fields will be integrated over $\hat r$.  But
this notation will allow us to focus on the symmetries for both types of
fields simultaneously.  In this notation, a term in the source spread out
over $S$ appears as
\begin{equation}
\cO\propto \prod_j\,\varphi^\dag_j(s,s,\hat r_j)
\label{ols}
\end{equation} 
This is not gauge invariant, but transforms as
\begin{equation}
\cO\to\cO\,\prod_j\exp\left[-i\,q_j\left.\Lambda(x_j)\right|_{t_j=r_j=s}
\right]
\label{multiply}
\end{equation}
To maintain gauge invariance we construct a compensating exponential on the light-shell
\begin{equation}
\,\exp\left(i\frac{e}{2\pi}\int\left(\sum_j\ell(\hat r,\hat r_j)\right)\
\partial_\mu A^\mu(x)
\,dS\right)
\label{li1}
\end{equation}
where $dS$ is our Lorentz covariant surface element on the small sphere.
\begin{equation}
dS=\theta(t)\,r\,\delta(r-s)\,\delta(r^2-t^2)\,d^4x
\label{spherelement}
\end{equation}
and assuming zero net charge
\begin{equation}
\ell(\hat r, r_j)=q_j\log(1-\hat r_j\cdot\hat
    r)
\label{logs}
\end{equation}
Putting all the pieces together, our gauge invariant
source on the light-shell, call it $\mathcal{S}$, is of the form 
{\renewcommand{\arraystretch}{3}
\begin{equation}
\begin{array}{c}
\displaystyle
C\lim_{s\to0}\int 
\exp\left(i\frac{e}{2\pi}\int\,\left(\sum_{j=1}^{m+n}
\ell(\hat r,\hat r_j)\right)\,
\partial_\mu A^\mu(x)
\,dS
\right)
\\ \displaystyle
\left(\prod_{j=1}^m\;r_j^{-1} \Phi^\dag_{j,E_j}(x_{j})\,\right)
\left(\prod_{j=m+1}^{m+n}r_j^{-2}\phi^{(\dag)}_j(x_{j})\,\right)
\left(\prod_{j=1}^{m+n}dS_j\right)
\end{array} 
\label{fullsource1}
\end{equation}
}where there are $n$ soft scalars $\phi^{(\dag)}$, $m$ LRE scalars $\Phi$,
and $dS_j$ refers to $dS$ with $x^\mu\to x_j^\mu$. Also, each LRE scalar
$\Phi$ will have an energy associated with it $E_i$, this is the energy
scaled out by the LRE expansion. Notice that there is a constant $C$, which
must be determined by matching.   

Assuming gauge transformations on the light-shell, our compensating exponential is unique. Also, one cannot
  help but notice the
 connection to the classical potential
  (\ref{classicalpotential}). 

\section{
Matching LRE fields}
\label{scalarmatching}

The simplest non-trivial matching to
consider is that of LRE scalars. For this, we will match the
amplitude of a source creating a one-particle state in the full theory to
the corresponding amplitude in the effective theory. Of course, for
  this source to be gauge invariant, the particle must be neutral. This
  allows us to focus on the LRE matching all by itself.  In the process, we
will define creation/annihilation operators in the EFT by relating them to
the familiar creation/annihilation operators in the full
theory. This construction can then be carried over trivially to
  interesting sources involving charged particles.
Let the matching condition be 
\begin{equation}
\langle \vec k|{\rm Full\, Source}|0\rangle\overset{\rm match}{=}
\langle \vec k|{\rm EFT\, Source}|0\rangle
\end{equation}
$\langle \vec k|$ is a one particle state for a scalar with momentum $k^\mu=(k,\vec k)$ as defined in the full theory. This matching will connect the position space of the effective theory to the momentum space of the full theory, as well as fix the coefficient of the effective theory source.   The full theory source is just $\phi(0)$. The EFT source for a high-energy particle, to leading order, has the form
\begin{equation}
c_1\int d \Omega_1\,s(\hat r_1)\Phi^\dag_{1,E_1}(s(\hat r_1),s(\hat r_1),z_1,\phi_1)
\end{equation}
where $c_1$ is the coefficient we will determine herein. The matching condition is then
\begin{equation}
\langle \vec k|\phi(0)|0\rangle\overset{\rm match}{=}\left\langle \vec k\left|c_1\int d \Omega_1\,s(\hat r_1)\Phi^\dag_{1,E_1}(s(\hat r_1),s(\hat r_1),z_1,\phi_1)\right|0\right\rangle
\label{MatchingCondition}
\end{equation}
The LHS is 1, and
the RHS of (\ref{MatchingCondition}) is
\begin{equation}\left\langle 0\left|\sqrt{2k}\,a_{k}\int d \Omega_1\,c_1\,s(\hat r_1)\Phi^\dag_{1,E_1}(s(\hat r_1),s(\hat r_1),z_1,\phi_1)\right|0\right\rangle
\label{sMatchRHS}
\end{equation}
Making the commutation of operators involved above well defined requires a few steps. First, define a full theory annihilation ($a_s$) operator in spherical coordinates by relating it to a standard full theory operator.
The familiar commutation relation is
\begin{equation}
\Bigl[a_p,a^\dagger_{p'}\Bigr]=(2\pi)^3\,\delta^{(3)}(\vec p-\vec p')
\end{equation}
which can be expressed in spherical coordinates as 
\begin{equation}
\Bigl[a_p,a^\dagger_{p'}\Bigr]=\frac{(2\pi)^3}{p^2}\,\delta(p-p')\,\delta(z-z')\,\delta(\phi-\phi')
\end{equation}
The spherical creation ($a_s$) and annihilation ($a_s^{\dag}$) operators we define by
\begin{equation}
\Bigl[a_p,a^\dagger_{p'}\Bigr]\equiv\frac{(2\pi)^2}{pp'}\,\Bigl[a_s(p,z,\phi),a_s^\dagger(p',z',\phi')\Bigr]
\end{equation}
Notice that we have
\begin{equation} \left(\frac{p}{2\pi}a_p\right)\left(\frac{p'}{2\pi}a^\dagger_{p'}\right)-\left(\frac{p'}{2\pi}a^\dagger_{p'}\right)\left(\frac{p}{2\pi}a_p\right)=a_{s}(p,z,\phi)a^\dag_{ s}(p',z',\phi')-a^\dag_{ s}(p',z',\phi')a_{s}(p,z,\phi)\end{equation}
so the relations between conventional and spherical creation and annihilation operators are 
\begin{equation} a_{s}(p,z,\phi)=\frac{p}{2\pi}a_p
\label{crelate}
\end{equation}
\begin{equation} a^\dag_{s}(p,z,\phi)=\frac{p}{2\pi}a^\dag_p
\label{annirelate}\end{equation}
 In the EFT we can write our fields in terms of creation/annihilation operators as 
\begin{equation}
\Phi^\dag(t,r,\hat r)=\int\,e^{ik(t-r)}\,\frac{1}{r}\,a^\dag_{LRE}(k,\hat
r)\,\frac{dk}{2\pi} 
\label{candaops}
\end{equation}
which is described in detail in appendix \ref{canquant}. Using the above two relations, (\ref{sMatchRHS}) becomes
\begin{equation}
=\int d \Omega_1\left\langle 0\left|\sqrt{2k}\;\frac{2\pi}{k}a_s(k,\hat k)\,c_1\int\,a^\dag_{LRE}(k_1,\hat r_1)\,\frac{dk_1}{2\pi}\right|0\right\rangle
\end{equation}
The final and crucial step is to notice that the commutation relations of
$a_s$ and $a_{LRE}$ (in appendix \ref{canquant}) look the same, but have
two important differences: $a_s$ involves angles in momentum space and the
energy involved is the full energy $k$, whereas  $a_{LRE}$ involves
angles in position space and
the residual momentum $k$. So we identify the 
angles in momentum space and position space and set 
\begin{equation}
a^\dag_s(k_{1}+k,\hat k)=a^\dag_{LRE}(k_{1},\hat r)
\end{equation}
This allows us to turn our $a^\dag_{LRE}$ into $a^\dag_s$. We find that the RHS of (\ref{MatchingCondition})
becomes 
\begin{equation}
=2\pi\sqrt{\frac{2}{k}}c_1
\end{equation}
So, we have $\ds c_1=\frac{1}{2\pi}\sqrt{\frac{k}{2}}$ and our full theory scalars relate to our LRE scalars as
\begin{equation}
\phi(0)=\frac{1}{2\pi}\sqrt{\frac{k}{2}}\int d \Omega_1\,\,s(\hat r_1)\,\Phi^\dag_{1,E_1}(s(\hat r_1),s(\hat r_1),\hat r_1)
\end{equation}
$c_1$ is the contribution from one LRE scalar to C in our general source
(\ref{fullsource1}), but we will have contributions from matching the other
LRE fields involved in the process as well. While this matching procedure
is fairly simple, it is essential for connecting the objects in LSET, which
are formulated in position space, to the momentum-space 
amplitudes one is accustomed to
calculating in the full theory. Relations for LRE photons operators,
analogous to those introduced here, will be described in the following
section. 

\section{Reproducing  full-theory results}

\subsection{Tree-level}
\label{treelevel}

We are now prepared to compare amplitudes in the full theory and effective theory. A relevant process to compare is one with the final state of an energetic photon, scalar (labelled by `-'), and anti-scalar (labelled by `+'). For this  comparison we will focus on the transverse component. In the full-theory, we have
\begin{equation}\langle\vec k\;\vec{p}_-\,\vec{p}_+|\phi^*(0)\phi(0)|0\rangle=\frac{e}{|k|}\left(\frac{\hat p_--\hat k(\hat p_-\cdot\hat k)}{1-\hat k\cdot\hat p_-}-\frac{\hat p_+-\hat k(\hat p_+\cdot\hat k)}{1-\hat k\cdot\hat p_+}\right)
\label{fullemission}
\end{equation}
On the LHS above, $\vec k$ refers to a transverse final photon state with
momentum $\vec k$. In the effective theory we want to do the calculation
with the same final state, but we now use our EFT source $\langle \vec
k\;\vec p_-\,\vec p_+|\mathcal{S}|0\rangle$. The dependence on scalar
factors disappears. After integrating by parts and making use of the
rescaling for LRE photons, 
we get
\begin{equation}\left\langle \vec k\left|\,\frac{-ie}{2\pi}\int 
\,\left(\cA^*_{Er}(x)\,(\partial_t+\hat r\cdot\vec\nabla)
+\frac{1}{\sqrt{2E}}\,\vec\cA^*_{E\perp}(x)\cdot\vec\nabla\right)
\left(\sum_{j=+,-}\ell(\hat r,\hat r_j)\right)\,dS\,
\right|0\right\rangle
\label{EFTemission2}
\end{equation}
Note that the exponential associated with the  LRE expansion has gone away because of the $\delta(r^2-t^2)$ in
  $dS$. Using the transformation that diagonalizes the kinetic energy (\ref{DiagTrans}) gives
{\renewcommand{\arraystretch}{3}
\begin{equation}
\begin{array}{c}
\displaystyle
=\left\langle \vec k\left|\frac{-ie}{2\pi}\int \,\Biggl( \left(\delta(x-x'){\untilde A}^*_{Er}(x')-{\vec{\untilde A}}_{E\perp}^{*}(x')\cdot\vec\nabla'
\,(\partial_t+ \vec\nabla\cdot\hat R)^{-1}
(x',x)/\sqrt{2E}\right)\,(\partial_t+\hat r\cdot\vec\nabla)
\right.\right.
\\ \displaystyle
\left.\left.+\delta(x-x')\frac{1}{\sqrt{2E}}\,{\vec{ \untilde A}}_{E\perp}^{*}(x')\cdot\vec\nabla'\Biggr)
\left(\sum_{j=+,-}\ell(\hat r,\hat r_j)\right)
dx'dS\,
\vphantom{\frac{1}{1}}\right|0\right\rangle
\end{array} 
\end{equation}
}
Where $(\partial_t +\vec\nabla\dot\hat R)^{-1}$ is given in (\ref{dtdrinverse}) and $\vec\del'$ involves derivatives with respect to $x'$. Since the final physical photon state is transverse, and the relevant propagator is diagonal, we can remove the term involving ${\untilde A}^*_{Er}$. Then simplifying and manipulating our differential operator gives
{\renewcommand{\arraystretch}{3}
\begin{equation}
\begin{array}{c}
\displaystyle
=\left\langle \vec k\left|\,\frac{-ie}{2\pi}\int \,\Biggl( \left(-{\vec{ \untilde A}}_{E\perp}^{*}(x')\cdot\vec\nabla'
\,(\partial_t+ \vec\nabla\cdot\hat R)^{-1}
(x',x)/\sqrt{2E}\right)\left(\left(\partial_t+\vec\nabla\cdot\hat r\right)-\frac{2}{r}\right)\right.\right.
\\ \displaystyle
\left.\left.+\delta(x-x')\frac{1}{\sqrt{2E}}\,{\vec{ \untilde A}}_{E\perp}^{*}(x')\cdot\vec\nabla'\Biggr)\left(\sum_{j=+,-}\ell(\hat r,\{\hat r_j\})\right)dx'dS\,
\vphantom{\frac{1}{1}}\right|0\right\rangle
\end{array} 
\end{equation}
}\begin{equation}=\left\langle \vec k\left|\,\frac{-ie}{\pi}\int \,\left({\vec{ \untilde A}}_{E\perp}^{*}(x')\cdot\vec\nabla'
\,(\partial_t+ \vec\nabla\cdot\hat R)^{-1}(x',x)
\frac{1}{\sqrt{2E}\,r}\right)\left(\sum_{j=+,-}\ell(\hat r,\{\hat r_j\})\right)dx'dS\,
\right|0\right\rangle
\label{EFTemission4}
\end{equation}
Now, as in (\ref{candaops}) for scalars, can write our transverse photon
field as
\begin{equation} 
 \vec {\untilde A}^*_{E\perp}(t',r',\hat r')=\int\,e^{ik(t'-r')}\,\frac{1}{r'}\,{\vec{ \untilde a}}_{E\perp}^{\dag}(k,\hat r')\,\frac{dk}{2\pi}
\end{equation}
 Using this in (\ref{EFTemission4}) gives  
 \begin{equation}=\left\langle 0\left|\sqrt{2|k|}\,\vec a_{k\perp}\,\frac{-ie}{\pi\sqrt{2E}}\int \,\left( e^{ik(t'-r')}\,\frac{1}{r'}{\vec{ \untilde a}}_{E\perp}^{\dag}\cdot\vec\nabla' \;(\partial_t+ \vec\nabla\cdot\hat R)^{-1}(x',x)\frac{1}{r}\right)\left(\sum_{j=+,-}\ell(\hat r,\{\hat r_j\})\right)\frac{dk}{2\pi}\,dx'dS\,
\right|0\right\rangle
\label{EFTemitOp}
\end{equation}
Again,  for the gauge fields' creation/annihilation operators, we can use
relations analogous to those introduced in the previous 
section for scalars,
\begin{equation}\vec a_{s}(p,z,\phi)=\frac{p}{2\pi}\vec a_p
\end{equation}
 Identifying
\begin{equation}
\vec a^\dag_s(k+E,\hat k)=\vec{\untilde{a}}^\dag_E(k,\hat r)
\end{equation}
allows us to have creation/annihilation operators with well-defined
commutation relations. Also, note that $E=|\vec k|$, and (\ref{EFTemitOp}) becomes 
\begin{equation}=\left\langle 0\left|\,\vec a_{s\perp}(E,\hat
      r_{k})\,\frac{-2ie}{E}\int \,\left( e^{ik(t'-r')}\frac{1}{r'}\,\vec
        a^\dag_{s\perp}(k+E,\hat r')\cdot\vec\nabla' \;(\partial_t+
        \vec\nabla\cdot\hat R)^{-1}(x',x)\frac{1}{r}\right)
\right.\right.
\end{equation}
\begin{equation}\left.\left.\times\left(\sum_{j=+,-}\ell(\hat r,\{\hat r_j\})\right)\frac{dk}{2\pi}\,dx'dS\, 
\right|0\right\rangle
\end{equation}
The relevant commutation relation is
\begin{equation}
\Bigl[\,\vec a_{s_\perp}(p,z,\phi)\,,\,\vec
a_{s\perp}^\dagger(p',z',\phi   ')\,\Bigr]
=2\pi\,P_\perp\,\delta(p-p')\,\delta(z-z')\,\delta(\phi-\phi')
\end{equation}
where $P_\perp$ is a projection operator for the perpendicular components. Using this we obtain
\begin{equation}=\frac{-2ie}{E}\int \,\left( \delta(\hat r_{k}-\hat r')e^{ik(t'-r')}\frac{1}{r'}\vec\nabla'_\perp \;(\partial_t+ \vec\nabla\cdot\hat R)^{-1}(x',x)\frac{1}{r}\right)\left(\sum_{j=+,-}\ell(\hat r,\{\hat r_j\})\right)dx'dS\,
\label{EFTemission5}
\end{equation}which involves
\begin{equation}
r \vec\nabla_\perp\left(\sum_{j=+,-}\ell(\hat r,\{\hat r_j\})\right)
= \sum_{j=+,-}q_j
\left(\frac{\hat r_j-\hat r(\hat r\cdot\hat r_j)}{1-\hat r\cdot\hat r_j}\right)
\label{nabla-log}
\end{equation}
 Using this along with integrating over $dx'$ and $dS$  in (\ref{EFTemission5}) gives
\begin{equation}
- \,\frac{ie}{E}\sum_{j=+,-}q_j\left(\frac{\hat r_j-\hat r_k(\hat r_k\cdot\hat r_j)}{1-\hat r_k\cdot\hat r_j}\right)\,
\label{EFTemissionResult}
\end{equation}
(\ref{EFTemissionResult}) has the same absolute magnitude as (\ref{fullemission}), confirming the structure of the effective theory.

\subsection{Double logs}
\label{doublelogs}

The simplest 1-loop correction to calculate is to the production of a
scalar/anti-scalar pair, having energies $E_+$ and $E_-$, respectively, and
with $\vec p_+\cdot\vec p_-=E_1E_2z_{+-}$,
\begin{equation}\left\langle \,\vec p_-\,\vec p_+\left|\mathcal{S}\right|0\right\rangle
\label{pairAmp}
\end{equation}
The
pointlike source can produce any energy and momentum, so the energies,
$E_{\pm}$ and directions of the scalars are unconstrained.
As discussed
above, the LRE expansion breaks the production process up into independent
sectors, and the matrix element (\ref{pairAmp}) picks out a particular
sector in which the only hard particles produced by the source are the
scalar and anti-scalar with the specified energies and angles.  
In the interactive theory, the radiative corrections
renormalize the source in this
sector to account for (for example) decreasing the amplitude for exclusive
production of only charged particles with no photons. Formally, the
renormalization factor is
\begin{equation}
\left.\int\,\exp\left(i\frac{e}{2\pi}\int\,
\left(\sum_{j=+,-}\ell(\hat r,\hat r_j)\right)\,
\partial_\mu A^\mu(x)
\,dS
\right)\,e^{i\int\mathcal{L}_{\rm LS}\,d^4x}\,[dA]
\right/
\int\,e^{i\int\mathcal{L}_{\rm LS}\,d^4x}\,[dA]
\label{functional2}
\end{equation}
Where $\mathcal{L}_{\rm LS}$ is the gauge Lagrangian in light-shell gauge. Notice that $x$ contributes the only non-angular dependence, but $dS$
involves delta functions that leave us with purely angular
dependence. This dependence solely on angles will persist for any
process to any order in LSET. 
Evaluating (\ref{functional2})
to order $e^2$ 
using the methodology introduced
in \cite{LSGprop} we arrive at 
\begin{equation}
=1-\frac{e^2}{64\pi^4}\sum_{j=+,-\atop k=+,-}\int\,\frac{\bigl(\hat r_1\times\hat r_j\bigr)\cdot
\bigl(\hat r_2\times\hat r_k\bigr)}
{(1-z_{1j})\,(1-z_{12})\,(1-z_{2k})}
\,d\Omega_1\,d\Omega_2
\label{formal}
\end{equation}
To obtain (\ref{formal}), we have manipulated distributions such that
the result is not well-defined without regularization. For $j\neq k$, the
only non-integrable singularity is at $\hat r_1=\hat r_2$.  
After regulating by taking $(1-z_{12})\to(1-z_{12}+\lambda)$ we find the
$j\neq k$ contribution 
\begin{equation}
\frac{e^{2}}{2\pi^2}\log\left(\lambda\right)\,
\log\left(1-z_{+-}\right)
\label{jneqk}
\end{equation}

For $j=k$ in (\ref{formal}), the calculation is much more delicate and
depends on the details of the angular cut-off around the $\hat r_+$ and
$\hat r_-$ directions.  But for physical consistency, the $\lambda$ and $\theta$
dependence must 
disappear as the hard emission becomes neutral.  
For example, if
$\theta_+\approx\theta_-\approx\theta\ll1$, the $\lambda$ dependence should
cancel as $\theta_{+-}\!\to\theta$, because in this limit, we have two
small, oppositely
charged and equal-sized jets sitting right on top of one another to the
level of accuracy to which we know their directions.
Furthermore we expect that the
$j=k$ contributions should depend on $\theta_+$ or $\theta_-$, but not
both, and should be symmetrical in the two.
We therefore expect the leading log 
$\lambda$ and  log $\theta$ dependence to have the form
\begin{equation}
-\frac{e^{2}}{2\pi^2}\log\left(\lambda\right)\,
\log\left(\frac{\theta_-\theta_+/2}{1-z_{+-}}\right)
\label{consist}
\end{equation}
Log squared $\theta$ dependence is not possible, because it cannot be
simultaneously dependent on $\theta_+$ and $\theta_-$ separately, 
symmetrical in $\theta_+\leftrightarrow\theta_-$ and cancel when 
$\theta_+\approx\theta_-\approx\theta$.

We now claim to have derived (\ref{consist}) from a
combination of explicit calculation and consistency, but we do still wish for a  better  understanding of the radiative correction calculation. 
If we now assume (without obvious justification for now) that
$\lambda$ is given by a product of the radial resolution, proportional to
$\sqrt{\cE/E_i}$, of the two LRE particles  
\begin{equation}
\lambda=\frac{\cE}{\sqrt{E_1E_{2}}}
\end{equation}
then (\ref{consist})  is
\begin{equation}
\frac{e^{2}}{2\pi^2}\log\left(\frac{\cE}{\sqrt{E_1E_{2}}}\right)\,
\log\left(\frac{\theta_-\theta_+}{1-z_{+-}}\right)
\end{equation}
This matches the double-logs from the corresponding  calculation in
0-flavor sQED. This result confirms that the leading contribution from LSET
can reproduce the physics of the full theory. Also, since our theory
automatically disallows soft and collinear emission, this is the only
double-log contribution that one should expect for the process at hand. In
the full theory, one must combine the 1-loop calculation with real soft and
collinear emission to achieve the same result.   
\section{Conclusion}
Here we have constructed light-shell effective theory in the simplest
possible case. New features essential to the construction include
light-shell gauge, the large radial energy expansion, and the LSET
source. After putting the pieces together, we showed that the effective
theory reproduces the 0-flavor scalar quantum electrodynamics tree-level
amplitude and (with appropriate choice of cut-offs) 1-loop double-logs.  

We aim to extend the effective theory approach introduced here in a few
ways. Our hope is to eventually reach a QCD on the light-shell. First we
will need to consider the effects of fermionic matter, higher order corrections,
and matter loops. 

The dream of the effective field theorist is that the theory will take over
--- that once the appropriate degrees of freedom and symmetries are
identified, everything else will fall into place.  No theory for general high-energy collisions achieves this so simply. SCET's derivation, for example, takes place in a particular subset of gauges \cite{SCETfactorOnshell:2013}. We are also not there yet
with LSET since, for example, we are still using the LSG gauge propagator off
the light shell as well as on.  However,
we find the results in this paper encouraging, and we hope they are the first
steps to understanding a light-shell effective field theory for high-energy
collisions.  

\section*{Acknowledgements}

We have benefited greatly from suggestions by Ben Grinstein, Aneesh Manohar, 
Matthew Schwartz, Randall Kelley, Yang Ting Chien, Ilya Feige, and
David Simmons-Duffin. Some of the initial work on this project by HG was
done at the Aspen Center for Physics.
He is grateful for the support of the Center
 and National Science Foundation grant \#1066293. GK thanks the National Science Foundation Graduate Research Fellowship Program for their support.
This research has also been supported at Harvard 
in part by the National Science Foundation
under grants PHY-0804450 and
and PHY-1067976. 

\appendix

\section{0-Flavor Scalar Quantum Electrodynamics}
\label{0-flavor}
We consider sQED with a gauge invariant source of scalars with various charges and
flavor quantum numbers at the origin of
space and time with a coefficient that goes to infinity as the number of
flavors goes to zero so that there is a well-defined limit in which we can
ignore matter loops.  The details do not matter very much, but for example 
we could have $k$ scalar fields $\phi_{j,\alpha_j}$ for $j=1$ to $k$
with charges $q_j$ and with 
the flavor label $\alpha_j$ running from $1$ to $n_j$ describing $n_j$
(identical) flavors.
Our source could then look like
\begin{equation}
\lambda\lim_{\{n\to0\}}\,\frac{1}{K(\{n\})}\sum_{\{\alpha\}}\,\kappa_{\alpha_1\cdots\alpha_k}\,
\phi^*_{1,\alpha_1}\cdots\phi^*_{k,\alpha_k}
\label{source}
\end{equation}
where 
\begin{equation}
\sum_{\{\alpha\}}\,|\kappa_{\alpha_1\cdots\alpha_k}|^2=K(\{n\})
\end{equation}
and $K(\{n\})\to0$ if any of the $n_j$s vanish.  
This describes the production of $k$ ``0-flavor'' scalars.
A trivial example is to
have two fields with opposite charges and the same flavor and 
take $\kappa=\delta_{\alpha_1\alpha_2}$ and $K=n$.
We go through this song and dance to assure the reader that we can
ignore matter loops in a mathematically consistent limit without otherwise putting
important restrictions on the physics.  The important thing about such a
source is that it produces the charged particles with charges $q_j$, and
henceforth we will drop the flavor indices and just ignore the matter loops. 

\section{LRE Canonical Quantization}
\label{canquant}
The
leading order Lagrangian for LRE scalars is
\begin{equation}
\cL=i\Phi_E^\dag\,\Bigl(
\partial_t+(\hat r\cdot\vec\nabla+\vec\nabla\cdot\hat r)/2
\Bigr)\Phi_E
\label{lrelagr0}
\end{equation}
If we take $\Phi_E$ as the canonical field, 
\begin{equation}
\Pi_E=i\Phi_E^\dag
\end{equation}
and we can write
\begin{equation}
\Phi_E(t,r,\hat r)=\int\,e^{-ik(t-r)}\,\frac{1}{r}\,a_{LRE}(k,\hat r)\,\frac{dk}{2\pi}
\label{psioneterm}
\end{equation}
with
\begin{equation}
\Bigl[a_{LRE}(k,z,\phi),a_{LRE}^\dag(k',z',\phi')\Bigr]=
2\pi\,\delta(k-k')\,\delta(z-z')\,\delta(\phi-\phi')
\label{coma}
\end{equation}
so that the $a_{LRE}$s are annihilation operators and the $a_{LRE}^\dag$s 
are creation operators.
We can use the properties of the creation and annihilation
operators to compute the propagator, (\ref{lreprop0}).

Here $\Phi_E$ is a linear combination of annihilation
operators.  The positive sign of
the $i$ in (\ref{lrelagr0}) is required to give the right sign of the commutation
relation (\ref{coma}).  This, in turn is related to having pulled
out a factor of $e^{-iE(t-r)}$ from $\phi$ and a factor of $e^{iE(t-r)}$
from $\phi^*$, so we have built in the fact that $\Phi$ annihilates the vacuum.

For the LRE analysis, it is crucial that $k$ is not a positive energy.  It is a
residual energy that can have either sign (of course it satisfies $|k|\ll\cE$).  Because of this, we only need the single term in
(\ref{psioneterm}) to give the canonical commutation relations.  It is not
possible to interpret (\ref{coma}) with only positive energies (at least,
not without introducing negative-norm states).

\section{Conventions}
\label{conventions}

We use the metric 
\begin{equation}
g^{\mu\nu}= \,\begin{pmatrix}
1&0\\
0& 
-I \\
\end{pmatrix}
\end{equation}
Conventions for gauge transformations are
\begin{equation}
D^\mu=\partial^\mu-ie\,A^\mu
\end{equation}
\begin{equation}
A^\mu\to A^\mu+\frac{1}{e}\,\partial^\mu\Lambda
\end{equation}
\begin{equation}
\phi(x)\to e^{i\,\Lambda(x)}\,\phi(x)
\end{equation}
Classical calculations have been done in Heaviside-Lorentz units.

\bibliography{shell}

\end{document}